\documentclass[aps,letter,twocolumn]{revtex4}
\usepackage{bm,graphicx,hhline}

\begin{document}

\title{Mixing of two-electron spin states in a semiconductor quantum dot}
\author{\c{S}.~C. B\u{a}descu}
\author{T. L. Reinecke}
\address{\it{Naval Research Laboratory,  Washington DC 20375}}
\date{\today}

\begin{abstract}
We show that the low lying spin states of two electrons in a
semiconductor quantum dot can be strongly mixed by
electron-electron asymmetric exchange. This mixing is generated by
the coupling of electron spin to its orbital motion and to the
relative orbital motion of the two electrons. The asymmetric
exchange can be as large as $50\%$ of the isotropic exchange, even
for cylindrical quantum dots. The resulting spin mixing
contributes to understanding spin dynamics in quantum dots,
including light polarization reversal.
\end{abstract}

 \maketitle

An electron spin in a semiconductor quantum dot (QD) is an
attractive qubit for quantum computing \cite{Loss98}: the spin in
the ground orbital state can have long coherence time
\cite{Kroutvar04}; a single qubit can be initialized or read
optically by transient electron-hole pair excitation giving a
negative trion $X^-$ \cite{Cortez02,Ware05}; and the manipulation
of the spin exchange between neighboring spins can be the basis
for two-qubit gates \cite{Loss98}. A detailed picture of
correlations between spins in QDs is essential for understanding
the spin dynamics. The dominant interaction between two electrons
({\it e-e}) is the Heisenberg-like spin-symmetric $J\hat{\bm
s}_1$$\cdot$$\hat{\bm s}_2$ (symmetric exchange), which conserves
the total spin $\hat{\bm S}$$=$$\hat{\bm s}_1$$+$$\hat{\bm s}_2$.
Additional spin-asymmetric {\it e-e} interactions (asymmetric
exchange) do not conserve $\hat{\bm S}$ and thus decrease the
fidelity of gate operations.

Among a number of recent experiments giving information about spin
dynamics are those involving an optical polarization reversal
\cite{Cortez02,Ware05}. For them, it has been suggested that this
effect results from spin flipping due to electron-hole ({\it e-h})
exchange in QDs with lateral asymmetry. However, these experiments
require strong spin mixing, inconsistent with {\it e-h} exchange
alone \cite{Ware05}.

Spin-orbit (s-o) interactions play a key role in understanding
mixing of spin states. They arise from effective magnetic fields
created by the orbital motion of electrons \cite{B-L-P}. Electrons
in QD ground states with dominant $s$ components have small
orbital angular momentum and thus small s-o coupling. A number of
experiments of interest involve excited electrons in excited
states of the QD. Linear combinations of nearly degenerate excited
states in a plane ({\it e.g.} $p_x$- and $p_y$-like) can give rise
to 2D orbital motion with an effective magnetic field
perpendicular to the plane, and thus to large s-o coupling. This
is analogous to the $\hat{\bm L}$$\cdot$$\hat{\bm S}$ coupling in
atoms \cite{L-L}. Thus, relatively symmetric QDs ({\it e.g.}
cylindrical) can have significant s-o effects, as we show here.

There are three sources of s-o coupling that lead to the mixing of
spin states. The largest two contributions arise from the ${\bm
k}$$\cdot$$\hat{\bm p}$ mixing between the conduction and valence
bands near the zone center, as described in the effective mass
approach \cite{BirPikus}. We derive them by treating the
potentials from the structure and from the {\it e-e} Coulomb
repulsion on the same footing with ${\bm k}$$\cdot$$\hat{\bm p}$
terms, using the Kane model \cite{Kane}. We have in mind QDs with
a strong confinement to a single state $\xi(z)$ along the growth
axis ${\bm e}_z$, and a weaker confinement in the transverse
directions, which give the electron states $\phi_i({\bm
r})$$=$$\xi(z)$$\varphi_i({\bm \rho})$.

The first contribution to the s-o coupling, $\hat{\bm h}^V$,
arises from the {\it structure potential} $V({\bm r})$ of the QD.
It gives a single-electron s-o coupling of the form
\cite{ConstantV}:
\begin{equation}
\hat{\bm h}^V\cdot\hat{\bm
s}=\gamma_s^V\left[\partial_zV\left(\hat{\bm p}^\bot\times\hat{\bm
s}^\bot\right)+\left(\partial_{\bm \rho}V\times\hat{\bm
p}^\bot\right)\hat{s}^z\right]{\bm e}_z\,,\label{h_R}
\end{equation}
where $p^z$ is not present due to the strong vertical confinement
(for a single state $\xi(z)$, $\langle \xi|p^z|\xi
\rangle$$=$$0$). The first term in Eq.(\ref{h_R}) is the usual
Rashba coupling $\gamma^V({\bm e}_z$$\times$$\hat{\bm p}^\bot)$,
where $\gamma^V$$=$$\gamma_s^V\langle\xi|\partial_z V|\xi\rangle$,
associated to asymmetry in the growth direction \cite{Rashba84}.
The second term is important for excited states whose main
components are inversion-asymmetric ($p$-like), where it gives the
dominant s-o coupling, independent of structure or bulk inversion
asymmetries. This term vanishes in the QD ground state, whose main
component is inversion-symmetric ($s$-like).

The second contribution, $\hat{\bm h}^C$, arises from the
interaction of each spin with the orbital motion of the other. We
have obtained it within a two-particle {\it ${\bm
k}$$\cdot$$\hat{\bm p}$} approach for electrons interacting
through the {\it Coulomb potential} $U_C({\bm r}_r)$
\cite{Badescu05,ConstantC}. For $k=$$1,2$ and ${\bm r}_r$$=$${\bm
r}_1$$-$${\bm r}_2$, we have:
\begin{equation}
\hat{\bm h}_k^C\cdot\hat{\bm s}_k=(-1)^k\gamma_s({\bm
\nabla}_{{\bm r}_r}U_C\times \hat{\bm p}_k)\cdot\hat{\bm
s}_k\,.\label{h_rels-o}
\end{equation}

The coupling $\hat{\bm h}^V$ from Eq.(\ref{h_R}) is analogous to
the Pauli s-o interaction, while $\hat{\bm h}^C$ from
Eq.(\ref{h_rels-o}) is analogous to the Breit-Pauli spin-relative
orbit coupling \cite{B-L-P}. The Pauli and Breit-Pauli couplings
in vacuum or in atoms are relativistically small, due to the large
energy gap $2m_0c^2$ between electron and positron bands, whereas
the present gap $E_g$ is smaller, giving larger s-o couplings.

There is also a smaller contribution, $\hat{\bm h}^\mathrm{B}$,
from the {\it Dresselhaus coupling} due to the lack of bulk
inversion symmetry \cite{Dresselhaus}. It arises from the mixing
of the conduction band with the remote upper bands and it gives a
single-particle s-o coupling in the form $\hat{\bm
h}^\mathrm{B}$$\cdot$$\hat{\bm
s}$=$\gamma^B_b\epsilon^{\alpha\beta\delta}\hat{p}_\alpha(\hat{p}_\beta^2$$-$$\hat{p}_\delta^2)\hat{s}_\alpha$,
with indices denoting crystal symmetry axis. In the QDs with
strong vertical confinement consider here $\langle
p^{z2}\rangle$$\gg$$\langle p^{\bot2}\rangle$, thus the effective
Dresselhaus coupling contains only the transverse components ${\bm
h}^{B,\bot}$$\simeq$$\gamma^B(\hat{p}_x,-\hat{p}_y)$, with
$\gamma^B$$=$$\gamma^B_{b}\langle\xi|p_z^2|\xi\rangle$.

We use a model of QDs \cite{ModelPotential} resembling those from
self-assembled growth \cite{SK} along crystal axis $[001]$. The
lateral potential $\mathcal{V}({\bm \rho})$ contains a part
$\mathcal{V}_s$ symmetric for the inversion ${\bm
\rho}$$\rightarrow$$-{\bm \rho}$, and it may also contain an
inversion-asymmetric part $\mathcal{V}_a$ \cite{AsymDots}. We take
the principal axes ${\bm e}_{x,y}$ of the QD to be along the
crystal axes $[110]$ and $[1\overline{1}0]$. To construct accurate
states we use a large basis set of harmonic oscillator
wavefunctions. The lateral sizes $a_x$,$a_y$ are given by the
curvature at the potential minimum \cite{Single-particle-Phi_i},
which is determined entirely by the symmetric part
$\mathcal{V}_s$. $\mathcal{V}_a$ contains
$\mathcal{V}_{ax}$~($\mathcal{V}_{ay}$), odd in $x$~($y$), and is
parametrized by $E_x$~($E_y$) \cite{ModelPotential}. The effective
lateral electric field in the ground state
$\langle\varphi_1|$$-\partial_{x,y}\mathcal{V}_a$$|\varphi_1\rangle\propto
E_{x,y}$ and vanishes for lateral inversion symmetry.

\begin{figure}[htbp]
\unitlength1cm
\begin{picture}(8.0,9.0) \includegraphics{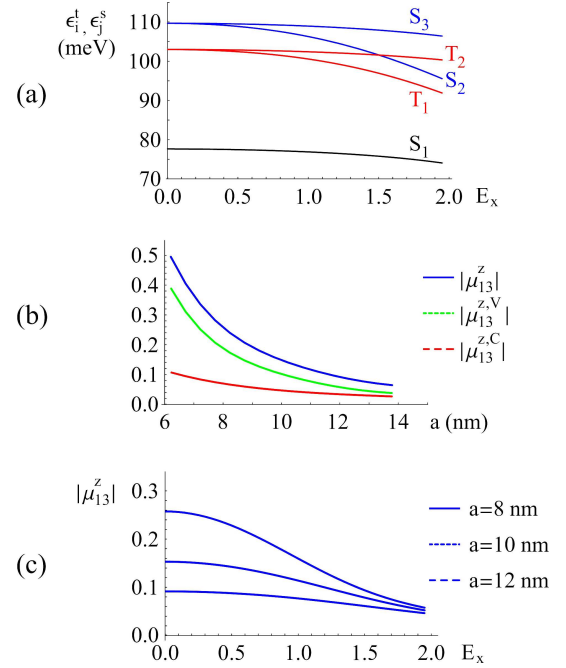}
\end{picture}
\caption{(Color online) QDs with $a_x$$=$$a_y$$=$$a$. {\bf (a)}
Two-electron energy levels in QDs with $a$$=$$10$~nm with one
plane of symmetry along ${\bm e}_x$ ($E_y$$=$$0$) vs the lateral
asymmetry parameter $E_x$. {\bf (b)} The asymmetric exchange
$|\mu_{13}^z|$ from Eq.(\ref{ratios}) and its components
$|\mu_{13}^{z,V}|$ (from s-o coupling), $|\mu_{13}^{z,C}|$ (from
spin-relative orbit coupling) vs the QD size $a$. {\bf (c)}
$|\mu_{13}^z|$ vs $E_x$$\neq$$0$ ($E_y$$=$$0$) for several QDs.}
\label{Asymmetry-Cases}
\end{figure}

First we consider the two-electron wavefunctions without s-o
coupling. They are obtained by diagonalizing hamiltonian $H_0$
that contains the Coulomb interaction $U_C$$=$$\frac{e^2}{\kappa
r_r}$ ($\kappa$ is the dielectric constant) with a band-mixing
correction $\gamma_c\delta({\bm r}_r)$ \cite{ConstantC} and the QD
potential $V({\bm r})$, in the basis of harmonic oscillator
wavefunctions. These basis functions separate into the symmetric
and antisymmetric sets $\{S^{(0)}_n\}$, $\{T^{(0)}_m\}$ by their
permutation symmetry \cite{Two-particle-basis}. In general, each
eigenstate of $H_0$ can be written in terms of functions having
definite $s$, $x$, $y$, $d$ symmetry, {\it e.g.}:
\begin{eqnarray}
T_1&=&T_1^x+E_xT_1^s+E_xE_yT_1^y+E_yT_1^d\,,\nonumber\\
T_2&=&T_2^y+E_yT_2^s+E_xE_yT_2^x+E_xT_2^d\,,\label{WF-comp}\\
S_2&=&S_2^x+E_xS_2^s+E_xE_yS_2^y+E_yS_2^d\,,\nonumber\\
S_3&=&S_3^y+E_yS_3^s+E_xE_yS_3^x+E_xS_3^d\,,\nonumber
\end{eqnarray}
where $T_1$~($T_2$) labels the lowest triplet with a larger
$x$~($y$) component, and $S_2$~($S_3$) labels the lowest singlet
with a larger $x$~($y$) component. The lowest lying states are
shown in Fig.\ref{Asymmetry-Cases}(a), where $E_y$$=$$0$. Higher
lying states not shown are $T_3$~($T_4$), which are the lowest $d$
($s$) symmetry triplets, and $S_4$ (the lowest $d$-symmetry
singlet). The isotropic part of the exchange for a triplet $T_i$
(with energy $\epsilon_i^t$) and a singlet $S_j$ (with energy
$\epsilon_j^s$) is given by
$J_{ij}$$=$$2($$\epsilon_i^t$$-$$\epsilon_j^s)/\hbar^2$. Here we
have chosen the energy splitting between the electron ground and
excited states to be in the range $20$$-$$45$~meV; this gives an
exchange splitting ($J_{13}$ between $T_1$ and $S_3$) of the order
$5$$-$$10$~meV, in the range of the experiments.

Next, the triplet-singlet mixing is generated by adding the s-o
terms ${\bm h}^V$, ${\bm h}^C$, ${\bm h}^B$ to $H_0$. These give a
hamiltonian composed of a spin-symmetric part $H_s$ that conserves
the total spin, and a spin-antisymmetric part $H_a$:
\begin{eqnarray}
H_s&=&H_{0}+\frac{1}{2}\left(\hat{\bm h}_1+\hat{\bm
h}_2+\gamma_s\partial_{{\bm \rho}_r}U_C\times \hat{\bm
p}_r^\bot\right)\cdot\hat{\bm S}\,,\label{Two_Particles-Diag}\\
H_a&=&\frac{1}{2}\left(\hat{\bm h}_1-\hat{{\bm
h}}_2+2\gamma_s\partial_{{\bm \rho}_r}U_C\times \hat{\bm
p}_c^\bot\right)\cdot\left(\hat{\bm s}_1-\hat{\bm
s}_2\right)\,,\nonumber
\end{eqnarray}
where $\hat{\bm h}_k$$=$$\hat{\bm h}_k^V$$+$$\hat{\bm h}_k^B$,
$\hat{\bm p}_r$$=$$\hat{\bm p}_1$$-$$\hat{\bm p}_2$, and $\hat{\bm
p}_{c}$$=$$(\hat{\bm p}_1$$+$$\hat{\bm p}_2)/2$. $H_a$ can be
written as:
\begin{eqnarray}
H_a&=&\sum_{i,j}{\bm \beta}_{ij}\cdot(\hat{\bm s}_1-\hat{\bm
s}_2)|T_i\rangle\langle S_j|+h.c.\label{H_st}\,,\\
{\bm \beta}_{ij}&=&\langle T_i|\hat{\bm
h}_1+\gamma_s\partial_{{\bm \rho}_r}U_C\times\hat{\bm
p}_c^\bot|S_j\rangle\nonumber\,,
\end{eqnarray}
where ${\bm \beta}_{ij}$ gives the asymmetric exchange. States of
different total spin $|\bm S|$ are coupled via the operator
$\hat{\bm s}_1$$-$$\hat{\bm s}_2$, which is equivalent to the
Dzyaloshinskii-Morya form $\frac{2i}{\hbar}$$(\hat{\bm
s}_1$$\times$$\hat{\bm s}_2)$ \cite{D-M}. The asymmetric exchange
can be written:
\begin{equation}
{\bm \beta}\cdot\left(\hat{\bm s}_1-\hat{\bm
s}_2\right)=\beta^z\left(\hat{s}^z_1-\hat{s}^z_2\right)+{\bm
\beta}^\bot\cdot\left(\hat{\bm s}^\bot_1-\hat{\bm
s}^\bot_2\right)\,.\label{IntroEq}
\end{equation}
The longitudinal component $\beta^z$ conserves the total spin
projection $S^z$, {\it i.e.}, it mixes singlets with triplets that
have $S^z$$=$$0$ ("longitudinal mixing"). This is equivalent to a
precession of the total spin around axis ${\bm e}_z$ ($\Delta
S^z$$=$$0$). The transverse components ${\bm \beta}^\bot$ mix
states with different total-spin projection ($|\Delta
S^z|$$\neq$$0$), equivalent with a total-spin precession around
in-plane axis ("transverse mixing").

It is convenient to group the operators from the matrix element
giving ${\bm \beta}_{ij}$ in Eq.(\ref{H_st}) into an axial vector
operator $\hat{\bm A}$$\equiv$$\hat{A}{\bm e}_z$ and two polar
vector operators $\hat{\bm P}$$\equiv$$\hat{P}{\bm e}_z$,
$\hat{\bm R}$$\equiv$$\hat{\bm R}^\bot$:
\begin{eqnarray}
\hat{\bm A}&=&2\gamma_s^V\left(\hat{\partial}_{{\bm\rho}_1}\mathcal{V}_{s}\times\hat{\partial}_{{\bm\rho}_1}\right)-2\gamma_s\left(\partial_{{\bm\rho}_r}U_C\times\hat{\partial}_{{\bm\rho}_c}\right)\,,\nonumber\\
\hat{\bm P}&=&2\gamma_s^V\left(\hat{\partial}_{{\bm\rho}_1}\mathcal{V}_{a}\times\hat{\partial}_{{\bm\rho}_1}\right)\label{A-P_components}\\
\hat{\bm R}&=&-2\gamma^V\left({\bm
e}_z\times\hat{\partial}_{{\bm\rho}_1}\right)+2\gamma^B\left({\bm
e}_x\hat{\partial}_{x_1}-{\bm e}_y\hat{\partial}_{y_1}
\right)\nonumber\,.
\end{eqnarray}
$\hat{\bm A}$ and $\hat{\bm P}$ include the vertical magnetic
field from the $2D$ motion in the nearly degenerate excited
states, and they generate $\beta^z$ in Eq.(\ref{IntroEq}).
$\hat{\bm R}$ arises from the Rashba and Dresselhaus terms, and it
generates ${\bm \beta}^\bot$.

\begin{widetext}
\squeezetable
\begin{table*}
\caption{The part of the longitudinal coupling $\beta_{ij}^z$
determined by $\hat{\bm A}$ [Eq.(\ref{A-P_components})]. The
matrix elements $A_{ij}^{\alpha\beta}$$=$$\langle
T_i^\alpha|\hat{A}|S_j^\beta\rangle$ are between wavefunction
components of definite symmetries [Eq.(\ref{WF-comp})].}
\begin{tabular}{c|c|c|c|c}
\hline
&$S_1$ ($\approx$~$s$-symmetry)&$S_2$ ($\approx$~$x$-symmetry) &$S_3$ ($\approx$~$y$-symmetry) & $S_4$ ($\approx$~$d$-symmetry)\\
\hline

$T_4$&&&&\framebox{\makebox[\totalheight]{$A_{44}^{sd}$}}~$+$\\
($\approx$~$s$) &
$E_xE_y[A_{41}^{xy}$$+$$A_{41}^{yx}$$+$$A_{41}^{sd}$$+$$A_{41}^{ds}]$
&
$E_y[A_{42}^{sd}$$+$$A_{42}^{yx}$$+$$E_x^2(A_{42}^{xy}$$+$$A_{42}^{ds})]$
&
$E_x[A_{43}^{sd}$$+$$A_{43}^{xy}$$+$$E_y^2(A_{43}^{xy}$$+$$A_{43}^{sd})]$
&
$E_x^2A_{44}^{xy}$$+$$E_y^2(A_{44}^{yx}$$+$$E_x^2A_{44}^{ds})$\\
\hhline{~|~|-|-|~} \hline \hhline{~|~|-|-|~}

$T_1$ && \vline \hfill \,& \hfill \framebox{\makebox[\totalheight]{$A_{13}^{xy}$}}$~+$ \hfill \vline &\\
($\approx$~$x$) &
$E_y[A_{11}^{xy}$$+$$A_{11}^{ds}$$+$$E_x^2(A_{11}^{sd}$$+$$A_{11}^{yx})]$
& \vline \hfill
$E_xE_y[A_{12}^{xy}$$+$$A_{12}^{yx}$$+$$A_{12}^{sd}$$+$$A_{12}^{ds}]$
\hfill&\hfill
$E_x^2A_{13}^{sd}$$+$$E_y^2(A_{13}^{ds}$$+$$E_x^2A_{13}^{yx})$
\hfill\vline &
$E_x[A_{14}^{xy}$$+$$A_{14}^{sd}$$+$$E_y^2(A_{14}^{ds}$$+$$A_{14}^{yx})]$
\\
\hline

$T_2$ && \vline \hfill \framebox{\makebox[\totalheight]{$A_{22}^{yx}$}}~$+$\hfill\, & \hfill \vline &\\
($\approx$~$y$) &
$E_x[A_{21}^{yx}$$+$$A_{21}^{ds}$$+$$E_y^2(A_{21}^{sd}$$+$$A_{21}^{xy})]$
& \vline \hfill
$E_x^2A_{22}^{ds}$$+$$E_y^2(A_{22}^{sd}$$+$$E_x^2A_{22}^{xy})$
\hfill &\hfill
$E_xE_y[A_{23}^{xy}$$+$$A_{23}^{yx}$$+$$A_{23}^{sd}$$+$$A_{23}^{ds}]$
\hfill\vline &
$E_y[A_{24}^{yx}$$+$$A_{24}^{sd}$$+$$E_x^2(A_{24}^{ds}$$+$$A_{24}^{xy})]$\\
 \hhline{~|~|-|-|~}\hline \hhline{~|~|-|-|~}

$T_3$&\framebox{\makebox[\totalheight]{$A_{31}^{ds}$}}~$+$&&&\\
($\approx$~$d$) &
$E_x^2A_{31}^{yx}$$+$$E_y^2(A_{31}^{xy}$$+$$E_x^2A_{31}^{sd})$ &
$E_x[A_{32}^{ds}$$+$$A_{32}^{yx}$$+$$E_y^2(A_{32}^{xy}$$+$$A_{32}^{sd})]$
&
$E_y[A_{33}^{ds}$$+$$A_{33}^{xy}$$+$$E_x^2(A_{33}^{yx}$$+$$A_{33}^{sd})]$
&
$E_xE_y[A_{34}^{xy}$$+$$A_{34}^{yx}$$+$$A_{34}^{sd}$$+$$A_{34}^{ds}]$\\
\hline
\end{tabular}\label{Table-A}

\caption{The part of the longitudinal coupling $\beta_{ij}^z$
determined by $\hat{\bm P}$ [Eq.(\ref{A-P_components})].
$P_{ij}^{\alpha\beta}$$=$$\langle
T_i^\alpha|\hat{P}|S_j^\beta\rangle$ are between wavefunction
components of definite symmetry [Eq.(\ref{WF-comp})].}\nonumber
\begin{tabular}{c|c|c|c|c}
\hline
&$S_1$ &$S_2$  &$S_3$ & $S_4$\\
\hline

$T_4$ &
$E_x[P^{sx}_{41}$$+$$P^{xs}_{41}$$+$$E_y^2(P^{yd}_{41}$$+$$P^{dy}_{41})]+$
&
$P^{sx}_{42}$$+$$E_x^2P^{xs}_{42}$$+$$E_y^2(P^{yd}_{42}$$+$$E_x^2P^{dy}_{42})+$
&
$P^{sy}_{43}$$+$$E_x^2P^{xd}_{43}$$+$$E_y^2(P^{zys}_{43}$$+$$E_x^2P^{dx}_{43})+$
&
$E_x[P^{sy}_{44}$$+$$P^{xd}_{44}$$+$$E_y^2(P^{ys}_{44}$$+$$P^{dx}_{44})]+$\\
&
      $E_y[P^{sy}_{41}$$+$$P^{ys}_{41}$$+$$E_x^2(P^{xd}_{41}$$+$$P^{ds}_{41})]$
      &
      $E_xE_y(P^{sy}_{42}$$+$$P^{xd}_{42}$$+$$P^{ys}_{42}$$+$$P^{dx}_{42})$
      &
      $E_xE_y(P^{sx}_{43}$$+$$P^{xs}_{43}$$+$$P^{yd}_{z43}$$+$$P^{dy}_{43})$
      &
      $E_y[P^{sx}_{44}$$+$$P^{yd}_{44}$$+$$E_y^2(P^{xs}_{44}$$+$$P^{dy}_{44})]$\\
\hhline{~|~|-|-|~}\hline\hhline{~|~|-|-|~}

$T_1$ &
$P^{xs}_{11}$$+$$E_x^2P^{sx}_{11}$$+$$E_y^2(P^{dy}_{11}$$+$$E_x^2P^{yd}_{11})+$
& \vline \hfill
$E_x[P^{xs}_{12}$$+$$P^{sx}_{12}$$+$$E_y^2(P^{sy}_{12}$$+$$P^{ys}_{12})]+$
&
$E_x[P^{xd}_{13}$$+$$P^{sy}_{13}$$+$$E_y^2(P^{sy}_{13}$$+$$P^{dx}_{13})]+$
\hfill\vline &
$P^{xd}_{14}$$+$$E_x^2P^{sy}_{14}$$+$$E_y^2(P^{dx}_{14}$$+$$E_x^2P^{ys}_{14})+$ \\
&
    $E_xE_y(P^{xd}_{11}$$+$$P^{sy}_{11}$$+$$P^{dx}_{11}$$+$$P^{ys}_{11})$
    & \vline \hfill
    $E_y[P^{xd}_{12}$$+$$P^{dx}_{12}$$+$$E_x^2(P^{sy}_{12}$$+$$P^{ys}_{12})]$
     \hfill& \hfill
    $E_y[P^{xs}_{13}$$+$$P^{dy}_{13}$$+$$E_x^2(P^{sx}_{13}$$+$$P^{yd}_{13})]$
    \hfill\vline &
    $E_xE_y(P^{xs}_{14}$$+$$P^{sx}_{14}$$+$$P^{dy}_{14}$$+$$P^{yd}_{14})$ \\
\hline

$T_2$ &
$P^{ys}_{21}$$+$$E_x^2P^{dx}_{21}$$+$$E_y^2(P^{sy}_{21}$$+$$E_x^2P^{ys}_{xd})+$
& \vline \hfill
$E_x[P^{ys}_{22}$$+$$P^{dx}_{22}$$+$$E_y^2(P^{sy}_{22}$$+$$P^{xd}_{22})]+$
&
$E_x[P^{yd}_{23}$$+$$P^{dy}_{23}$$+$$E_y^2(P^{sx}_{23}$$+$$P^{xs}_{23})]+$
\hfill\vline&
$P^{yd}_{24}$$+$$E_x^2P^{dy}_{24}$$+$$E_y^2(P^{sx}_{24}$$+$$E_x^2P^{xs}_{24})+$ \\
&
    $E_xE_y(P^{yd}_{21}$$+$$P^{dy}_{21}$$+$$P^{sx}_{21}$$+$$P^{xs}_{21})$
    & \vline \hfill
    $E_y[P^{yd}_{22}$$+$$P^{sx}_{22}$$+$$E_x^2(P^{dy}_{22}$$+$$P^{xs}_{22})]$
     \hfill& \hfill
    $E_y[P^{ys}_{23}$$+$$P^{sy}_{23}$$+$$E_x^2(P^{dx}_{23}$$+$$P^{xd}_{23})]$
    \hfill\vline&
    $E_xE_y(P^{ys}_{24}$$+$$P^{dx}_{24}$$+$$P^{sy}_{24}$$+$$P^{xd}_{24})$ \\
\hhline{~|~|-|-|~}\hline\hhline{~|~|-|-|~}

$T_3$ &
$E_x[P^{dx}_{31}$$+$$P^{ys}_{31}$$+$$E_y^2(P^{xd}_{31}$$+$$P^{sy}_{31})]+$
&
$P^{dx}_{32}$$+$$E_x^2P^{ys}_{32}$$+$$E_y^2(P^{xd}_{32}$$+$$E_x^2P^{sy}_{32})+$
&
$P^{dy}_{33}$$+$$E_x^2P^{yd}_{33}$$+$$E_y^2(P^{xs}_{33}$$+$$E_x^2P^{sx}_{33})+$
&
$E_x[P^{dy}_{34}$$+$$P^{yd}_{34}$$+$$E_y^2(P^{xs}_{34}$$+$$P^{sx}_{34})]+$ \\
&
    $E_y[P^{dy}_{31}$$+$$P^{xs}_{31}$$+$$E_x^2(P^{yd}_{31}$$+$$P^{sx}_{31})]$
    &
    $E_xE_y(P^{dy}_{32}$$+$$P^{yd}_{32}$$+$$P^{xs}_{32}$$+$$P^{sx}_{32})$
    &
    $E_xE_y(P^{dx}_{33}$$+$$P^{ys}_{33}$$+$$P^{xd}_{33}$$+$$P^{sy}_{33})$
    &
    $E_y[P^{dx}_{34}$$+$$P^{xd}_{34}$$+$$E_x^2(P^{ys}_{34}$$+$$P^{sy}_{34})]$ \\
\hline
\end{tabular}\label{Table-Pd}
\end{table*}
\end{widetext}
\clearpage

Table \ref{Table-A} gives the matrix elements between $T_i$
($i$$=$$1,4$), and $S_j$ ($j$$=$$1,4$), from the spin mixing
operator $\hat{\bm A}$ in Eq.(\ref{A-P_components}). The states
are characterized by the symmetry of their dominant wavefunction
components, {\it e.g.} $S_2$ $\approx x$-symmetry. Table
\ref{Table-Pd} gives corresponding results from $P$. The terms in
small boxes in Table \ref{Table-A} are dominant and are
independent of lateral asymmetries. All the other terms in Tables
I and II are non-zero only for cases of lateral asymmetry. The
central $2\times 2$ block highlighted is of interest for the
dynamics of $X^-$ in the "$p$" shell \cite{Cortez02,Ware05}.

The matrix elements in Tables I and II can be understood by
writing the operators in the basis $\{T^{(0)}_m,{S^{(0)}_n\}}$:
$\hat{\bm A}$$+$$\hat{\bm
P}$$=$$\sum_{m,n}\left(A_{mn}+P_{mn}\right){\bm
e}_z|T^{(0)}_m\rangle\langle S^{(0)}_n|+h.c.$. The matrix elements
$A_{ij}^{\alpha\beta}$ and $P_{ij}^{\alpha\beta}$ in the tables
are sums of the matrix elements $A_{mn}$ and $P_{mn}$ with the
same symmetry. From Eq.(\ref{A-P_components}), it is seen that
$A_{mn}$ is nonzero only for $|T^{(0)}_m\rangle\langle S^{(0)}_n|$
odd both in $x$ and in $y$, thus $\hat{\bm A}$ can produce
longitudinal mixing $\beta^z_{ij}$ between two-electron
eigenstates $T_i$ and $S_j$ if one of these contains a
$x$~$(s)$-symmetry component, and the other has a
$y$~($d$)-symmetry part [Table \ref{Table-A}]. $P_{mn}$ is
non-zero only for QD asymmetries ($\mathcal{V}_a$$\neq$$0$) and
for $|T^{(0)}_m\rangle\langle S^{(0)}_n|$ odd either only in $x$
or only in $y$. Thus, $\hat{\bm P}$ contributes to the
longitudinal spin mixing $\beta^z_{ij}$ between $T_i$ and $S_j$ if
one of them has a $s$ or $d$ component and the other has a $x$ or
$y$ component [Table \ref{Table-Pd}]. $\hat{\bm R}$ can be written
as $\hat{\bm R}$$=$$\sum_{m,n}{\bm
R}^\bot_{mn}|T^{(0)}_m\rangle\langle S^{(0)}_n|+h.c.$ Results for
the matrix elements of ${\bm R}^\bot_{mn}$ are not given
explicitly here. They require QD lateral asymmetry and are nonzero
for $|T^{(0)}_m\rangle\langle S^{(0)}_n|$ odd in one of $x$ or
$y$. They can give transverse spin mixing ${\bm \beta}^\bot_{ij}$
of states with different $z$ spin projection. The degree of
triplet-singlet mixing is given by the ratio of the asymmetric to
the symmetric exchange:
\begin{equation}
\mu_{ij}^{z}=\hbar^{-1}\beta_{ij}^z/J_{ij}\,\,\,,\,\,\,{\bm
\mu}_{ij}^\bot\,=\,\hbar^{-1}{\bm
\beta}_{ij}^\bot/J_{ij}\,.\label{ratios}
\end{equation}

We now consider QDs with different asymmetries and consider the
longitudinal spin mixing $\mu^z$ from them. This mixing does not
have contributions from the Dresselhaus and Rashba couplings.

\begin{figure}[htbp]
\unitlength1cm
\begin{picture}(8.0,3.0)
\includegraphics{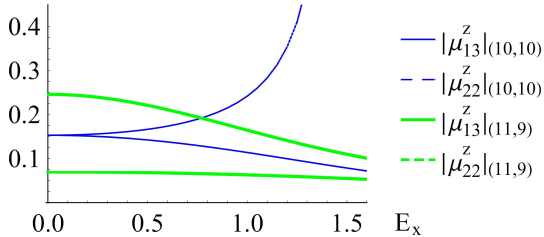}
\end{picture}
\caption{(Color online) QDs with $a_x$$\neq$$a_y$. Asymmetric
exchange $\mu^z$ in QDs with $a_x$$=$$11$~nm and $a_y$$=$$9$~nm
compared to the longitudinal coupling in QDs with
$a_x$$=$$a_y$$=$$10$~nm.}\label{EllipticQD}
\end{figure}

{\it i. QDs with lateral inversion symmetry}. For them
$E_x$$=$$E_y$$=$$0$. Examples are shown in
Fig.\ref{Asymmetry-Cases}(b) and by the $E_x$$=$$0$ points in
Figs.\ref{Asymmetry-Cases}(a,c) and Fig.\ref{EllipticQD}. In such
QDs, the two-electron states $T_i$, $S_j$ have well-defined
symmetries. The spin-mixing is due only to $\hat{\bm A}$, on the
second diagonal (in small boxes) in Table \ref{Table-A}. "Pure"
states of $x$~($y$)-symmetry such as $T_1$~($T_2$) couple only to
"pure" states of $y$~($x$)-symmetry such as $S_3$~($S_2$). The
first order longitudinal spin mixing of $T_1$~($T_2$) is from
$S_3$~($S_2$), which is the closest in energy. $T_3$ (the lowest
$d$-symmetry triplet) couples by $\hat{\bm A}$ to $s$-symmetry
singlets like $S_1$. $T_4$ (the lowest $s$-symmetry triplet)
couples by $\hat{\bm A}$ to $d$-symmetry singlets such as $S_4$.

From Fig.\ref{Asymmetry-Cases}(b) and Fig.\ref{EllipticQD} (at
$E_x$$=$$0$) we can see that the asymmetric exchange can be a
substantial fraction of the symmetric exchange (up to
$\approx50\%$). Fig.\ref{Asymmetry-Cases}(b) shows that the
asymmetric exchange is smaller for larger QDs, which results from
larger orbits giving smaller effective magnetic fields in the s-o
coupling. In this case $\beta_{22}^z$$=$$-\beta_{13}^z$ because of
degeneracy. The orbital momentum $\hat{L}_z$ eigenstates
$\frac{1}{\sqrt{2}}(S_2$$\pm$$iS_3)$ are strongly coupled to
$\frac{1}{\sqrt{2}}(T_1$$\pm$$iT_2)$ and obey $\Delta L_z$$=$$0$.
From Fig. \ref{EllipticQD} we see that the asymmetric exchange
decreases as the degeneracy of the first two excited states is
removed by different $a_x$ and $a_y$. In this case $L_z$ is not
conserved. The stronger confinement along ${\bm e}_y$
($a_x$$>$$a_y$) leads to $J_{13}$$>$$J_{22}$, and thus to
$|\mu_{13}^z|$$<$$|\mu_{22}^z|$.

\begin{figure}[htbp]
\unitlength1cm
\begin{picture}(8.5,2.8)
 \includegraphics{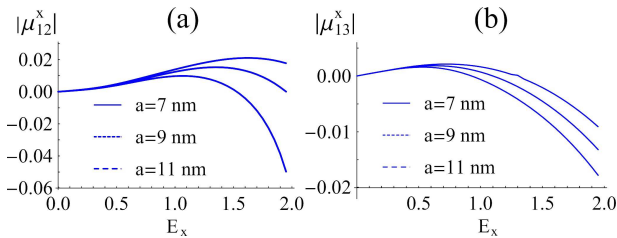}
\end{picture}
\caption{(Color online) Mixing of states with different spin
projection $S^z$ in QDs with $a_x$$=$$a_y$ and with a plane of
symmetry ($E_y$$=$$0$, $E_x$$\neq$$0$): ({\bf a}) $\mu^{x}_{12}$
for the mixing of $T_1$ and $S_2$. ({\bf b}) $\mu^{x}_{13}$ for
the mixing of $T_1$ and $S_3$.} \label{Figure-Relaxation}
\end{figure}

{\it ii. QDs with a single vertical plane of reflection}. For this
case $E_x$$\neq$$0$ and $E_y$$=$$0$. This gives more non-zero
matrix elements in Tables \ref{Table-A} and \ref{Table-Pd}, {\it
e.g.} now $T_4$ is mixed with $S_3$ as well as with $S_4$. This
case is illustrated in Fig.\ref{Asymmetry-Cases}(a,c) and in
Fig.\ref{EllipticQD}. $\mu^z$ for the lowest triplet is seen to
decrease with increasing $E_x$. For these cases, the terms
proportional to $E_x$ and $E_x^2$ in Table \ref{Table-A} and also
the terms from $P$$=$$2\widetilde{\gamma}^V$$(\partial_{{\bm
\rho}_1}$$\mathcal{V}_{ax}$$\times$$\hat{\partial}_{{\bm
\rho}_1})_{mn}^z\propto E_x$ from Table \ref{Table-Pd} are
nonzero, and they tend to cancel partially the larger terms in the
boxes in Table \ref{Table-A}. For some triplet-singlet pairs, such
as $S_3$ and $T_1$, the symmetric exchange becomes larger and thus
their mixing decreases. Other singlet-triplet pairs can be
degenerate, such as $T_2$ and $S_2$ in
Fig.\ref{Asymmetry-Cases}(a) at $E_x\approx1.5$; then nonzero
$\beta^z_{22}$ leads to strong singlet-triplet mixing
[Fig.\ref{EllipticQD}]. For this case, $L_z$ is not conserved.
Triplets with $\langle\hat{L}_z\rangle$$\approx$$\pm\hbar$ can be
coupled to singlets that have
$\langle\hat{L}_z\rangle$$\approx$$\mp\hbar$.

{\it iii. QDs with no vertical plane of reflection}. For this case
$E_x$$\neq$$0$ and $E_y$$\neq$$0$. Then all states in Tables
\ref{Table-A} and \ref{Table-Pd} are mixed, and the degree of
longitudinal spin mixing can be larger than in the previous cases.

In addition to the longitudinal spin-mixing described above, there
is also mixing that changes the spin projection $S^z$ (transverse
mixing ${\bm \mu}^\bot$). This arises exclusively from the
Dresselhaus and Rashba couplings, which give ${\bm R}$ in
Eq.(\ref{A-P_components}). For QDs with lateral inversion
symmetry, ${\bm R}$ mixes states which typically differ by the
single-particle energy splitting, {\it e.g.} $T_1$ with $S_1$ and
$S_4$ {\it etc.}. For them the mixing from $\hat{\bm R}$ is small,
due to large $J_{11}$ and $J_{14}$. For QDs with only one vertical
plane of reflection, $\hat{\bm R}$ mixes $T_1$ with $S_2$ or
$S_3$, which are closer in energy and therefore give larger
mixing. We show in Fig.\ref{Figure-Relaxation} this transverse
spin-mixing for $T_1$ and $S_2$ and for $T_1$ and $S_3$. This
mixing occurs only for non-zero asymmetric potential
($E_x$$\neq$$0$). It is generally smaller than the longitudinal
spin mixing discussed earlier, but it can become appreciable for
large asymmetries, and it is larger in smaller QDs.

In recent experiments on light polarization reversal in QDs after
excited state pumping \cite{Cortez02,Ware05}, the separation
between excited $p_x$ and $p_y$-like states is small and cannot be
resolved. One picture of this effect \cite{Cortez02,Ware05} is
that it involves mixing of electron triplets and singlets with
simultaneous spin flips of an electron and of the hole caused by
axially-asymmetric {\it e-h} exchange. For this mechanism,
however, the experimental results require large QD asymmetries
\cite{Ware05}. The additional mixing of triplets and singlets by
asymmetric {\it e-e} exchange here gives consistency with
experiments using more realistic QD potentials.

In addition, the present results for triplet-singlet mixing
provide an alternate process for the light polarization reversal.
The angular momentum from the light can be given to the orbital
motion of excited-state electrons. Then the triplet-singlet
coupling given here can mix two-electron states that differ in
their {\it orbital} angular momentum by $2\hbar$, leading to
reversed light polarization. This is given for example in QDs with
unequal lateral sizes even in the inversion-symmetric case ({\it
i}) above.

We are grateful for discussions with Y. Lyanda-Geller, B. V.
Shanabrook, D. Gammon and M. E. Ware. This work was supported in
part by the ONR and by DARPA.


\begin{thebibliography}{200}

\bibitem{Loss98} D. Loss, D. P. DiVincenzo, Phys. Rev. A {\bf 57}, 120 (1998)

\bibitem{Kroutvar04}M. Kroutvar {\it et al.}, Nature {\bf 432} (7013), 81
(2004)

\bibitem{Cortez02} S. Cortez {\it et al.}, Phys. Rev. Lett. {\bf 89}(20),
207401 (2002)

\bibitem{Ware05} M. E. Ware {\it et al.}, Phys. Rev. Lett. {\bf 95}(17), 177403
(2005)

\bibitem{B-L-P} V. B. Berestetskii, E. M. Lifsitz, L. P. Pitaevskii, {\it Quantum Electrodynamics}, (Butterworth-Heinemann, 1998)

\bibitem{L-L} L. D. Landau and E. M. Lifshitz, {\it Quantum Mechanics}, (Butterworth-Heinemann, 1998)

\bibitem{BirPikus}G. L. Bir and G. E. Pikus, {\it Symmetry and Strain-induced Effects in Semiconductors}, (John Wiley $\&$
Sons, 1974)

\bibitem{Kane} The Kane model \cite{BirPikus} uses parameters: band gap $E_g$, energy of the split-off band $\Delta$, and matrix
element $P$ of operator $\hbar\hat{p}_x/m_0$ between the Bloch
states of the conduction and valence bands.

\bibitem{ConstantV} $\gamma_s^V$ comes from the combination
of the ${\bm k}$$\cdot$$\hat{\bm p}$ band mixing with the
potential for holes $V_h$$=$$-c_{h} V$ ($0$$<$$c_{h}$$<$$1$),
which gives $\gamma_s^V$$=$$-\frac{c_h}{\hbar^2}\frac{2P^2}{3
E_g^2}\frac{\Delta (2 E_g +\Delta)}{(E_g +\Delta)^2}$. The
procedure is the same as that for treatment of the Coulomb
interaction \cite{Badescu05,ConstantC}

\bibitem{Rashba84} Yu. L. Bychkov, E. I. Rashba, JETP Lett. {\bf
39}, 78 (1984)

\bibitem{Badescu05}\c{S}.~C. B\u{a}descu {\it et al.}, Phys. Rev. B {\bf 72}, 161304(R) (2005)

\bibitem{ConstantC} $\gamma_s$ comes from the combination
of the ${\bm k}$$\cdot$$\hat{\bm p}$ terms with the {\it e-e}
Coulomb potential, which gives
$\gamma_s$$=$$\frac{1}{\hbar^2}\frac{2P^2}{3 E_g^2}\frac{\Delta (2
E_g +\Delta)}{(E_g +\Delta)^2}$ \cite{Badescu05}. This also gives
to a spin-independent correction to the energy,
$\gamma_c$$\delta({\bm r})$, where $\gamma_c$$=$$2\pi
\frac{e^2}{\varepsilon}\frac{2P^2}{3 E_g^2}\frac{(E_g
+\Delta)^2+E_g^2}{(E_g +\Delta)^2}$.

\bibitem{Dresselhaus} G. Dresselhaus, Phys. Rev. {\bf 100}, 580 (1955).

\bibitem{ModelPotential}We consider dots of height $W$, with confining potential $V({\bm
r})$$=$$-$$U_0\theta(|z$$-$$W/2|)(1$$+$$E_z
z)\widetilde{\mathcal{V}}({\bm \rho})$, where $U_0$ is the
conduction band offset and $\theta$ is the step function.  The
vertical function $\xi (z)$ is the solution of
$V_z(z)$$=$$-$$U_0\theta(|z$$-$$W/2|)(1$$+$$E_z z)$, and
$\varphi_i$ are solutions of the lateral potential
$\mathcal{V}({\bm \rho})$$=$$-$$U_0 \widetilde{\mathcal{V}}({\bm
\rho})$. For the symmetric part of the potential we use the
Gaussian form $\mathcal{V}_s({\bm \rho})$$=$$-U_0
e^{-\left(x/A_x\right)^2-\left(y/A_y\right)^2}$, and for the
antisymmetric part we use $\mathcal{V}_a({\bm \rho})$$=$$
\mathcal{V}_s({\bm
\rho})\left(E_x\left(\frac{x}{A_x}\right)^3+E_y\left(\frac{y}{A_y}\right)^3\right)$.
$E_{x,y}$ are perturbations to $\mathcal{V}_s$ and characterize
the lateral inversion asymmetry of the system. The energy scale of
the lowest electron states is controlled by $A_{x,y}$ and is
chosen comparable to the experiment. Here we take $W$$=$$4$~nm and
use InAs/GaAs parameters with a band offset $U_0$$=$$0.6$~eV.

\bibitem{SK} W. Seifert {\it et al.}, Prog. Cryst. Growth Ch. {\bf33}, 423 (1996)

\bibitem{AsymDots} R. Krebs {\it et al.}, J. Cryst. Growth {\bf 251}, 742 (2003)

\bibitem{Single-particle-Phi_i}The single-particle basis is reducible to four subspaces:
$\{|\mathrm{s}\rangle\}$ even in $x$ and $y$,
$\{|\mathrm{x}\rangle\}$ odd in $x$, $\{|\mathrm{y}\rangle\}$ odd
in $y$, and $\{|\mathrm{d}\rangle\}$ odd in $x$ and in $y$. If the
system has inversion symmetry, these provide four independent
subspaces for the eigenstates.

\bibitem{Two-particle-basis}The two-particle basis is reducible to four subspaces, respectivelly of
$s$, $x$, $y$, and $d$-symmetry. Symmetric combinations
$S^{(0)}_n$ form the singlet basis
$\{\sigma_{\mathrm{ss'}},\sigma_{\mathrm{xx'}},\sigma_{\mathrm{yy'}},\sigma_{\mathrm{dd'}}\}$$\oplus$$\{\sigma_{\mathrm{sx}},\sigma_{\mathrm{yd}}\}$$\oplus$$\{\sigma_{\mathrm{sy}},\sigma_{\mathrm{xd}}\}$$\oplus$$\{\sigma_{\mathrm{xy}},\sigma_{\mathrm{sd}}\}$;
antisymmetric combinations $T^{(0)}_m$ form the triplet basis
$\{\tau_{\mathrm{ss'}},\tau_{\mathrm{xx'}},\tau_{\mathrm{yy'}},\tau_{\mathrm{dd'}}\}$$\oplus$$\{\tau_{\mathrm{sx}},\tau_{\mathrm{yd}}\}$$\oplus$$\{\tau_{\mathrm{sy}},\tau_{\mathrm{xd}}\}$$\oplus$$\{\tau_{\mathrm{xy}},\tau_{\mathrm{sd}}\}$.

\bibitem{D-M}I. E. Dzyaloshinskii, Phys. Chem. Solids 4, 241 (1958); T.
Morya, Phys. Rev. 120, 91 (1960)


\end{thebibliography}
\end{document}